\documentclass[article,onecolumn]{IEEEtran}

\usepackage{amsthm,amsmath,amssymb,cite}
\usepackage{eucal}
\usepackage{xifthen}
\usepackage{mathtools}
\usepackage{enumerate}
\usepackage{microtype}
\usepackage{xspace}
\usepackage{bm}
\usepackage{lastpage}
\usepackage[center]{caption}
\usepackage{xcolor}
\allowdisplaybreaks
\usepackage{enumitem}

\newcommand{\mset}[1]{\widetilde{#1}_{\mathcal{L}}}
\newcommand{\ceiling}[1]{\lceil #1\rceil}

\newtheorem{theorem}{Theorem}

\newtheorem{example}{Example}

\DeclareMathAlphabet{\mathcal}{OMS}{cmsy}{m}{n}

\usepackage{graphicx}
\usepackage[font=small]{caption}
\usepackage{subcaption}

\begin{document}

\title{On the Capacity of Secure Distributed Matrix Multiplication}

\author{
Wei-Ting~Chang \qquad Ravi~Tandon\\ 
Department of Electrical and Computer Engineering\\
University of Arizona, Tucson, AZ, USA\\
E-mail: \{\textit{wchang, tandonr}\}@email.arizona.edu
}


\maketitle

\begin{abstract}
Matrix multiplication is one of the key operations in various engineering applications. Outsourcing large-scale matrix multiplication tasks to multiple distributed servers or cloud is desirable to speed up computation. However, security becomes an issue when these servers are untrustworthy. In this paper, we study the problem of secure distributed matrix multiplication from distributed untrustworthy servers. This problem falls in the category of secure function computation and has received significant attention in the cryptography community. However, the fundamental limits of information-theoretically secure matrix multiplication remain an open problem. We focus on information-theoretically secure distributed matrix multiplication with the goal of characterizing the minimum communication overhead. The capacity of secure matrix multiplication is defined as the maximum possible ratio of the desired information and the total communication received from $N$ distributed servers. In particular, we study the following two models where we want to multiply two matrices $A\in\mathbb{F}^{m\times n}$ and $B\in\mathbb{F}^{n\times p}$: $(a)$ one-sided secure matrix multiplication with $\ell$ colluding servers, in which $B$ is a public matrix available at all servers and $A$ is a private matrix. $(b)$ fully secure matrix multiplication with $\ell$ colluding servers, in which both $A$ and $B$ are private matrices. The goal is to securely multiply $A$ and $B$ when any $\ell$ servers can collude. For model $(a)$, we characterize the capacity as $C_{\text{one-sided}}^{(\ell)}=(N-\ell)/N$ by providing a secure matrix multiplication scheme and a matching converse. For model $(b)$, we propose a novel scheme that lower bounds the capacity, i.e., $C_{\text{fully}}^{(\ell)}\geq (\lceil \sqrt{N}-\ell \rceil)^2/(\lceil \sqrt{N}-\ell \rceil+\ell)^2$.

\end{abstract}


\section{Introduction}
\footnote{This work was supported by the NSF Grant CAREER-1651492.}
In the era of Big Data, performing computationally intensive operations on a local machine becomes challenging and inefficient. Relying on powerful distributed servers is desirable for improving efficiency. As clients, users can upload their data onto servers, and let servers perform computationally expensive tasks for them. However, if the servers are untrustworthy and the data contain sensitive information, it raises security concerns. Therefore, designing algorithms to take advantage of the powerful untrusted servers while keeping them from learning anything about input data is of significant interest.

Cryptography community has looked at this problem under the secure multi-party computation framework, also known as secure function evaluation. In a secure function evaluation problem, parties want to jointly compute a function without revealing their respective input to other parties. For example, Alice, who has input $x$, wants to compute $\mathit{f}(x,y)$ without leaking $x$ to Bob, who has input $y$, where $\mathit{f}$ is some function they want to compute jointly. Similarly, Bob does not want to reveal $y$ to Alice. Alice and Bob should not learn anything about each other's input from the result of the computation, either. Some previous works include secure two-party computation \cite{Yao1982} which proposed using one-way functions to achieve security, and secure multi-party computation \cite{Multiparty1988,SMPC2000} to name a few. A class of encryption schemes called Fully Homomorphic Encryption guarantees that any unencrypted items, including the inputs, any intermediate values and the outputs will not be leaked to unintended party. Naturally, it is often used as a solution to secure function evaluation problems and other types of security problems \cite{SFEHomo2009,Homo2010Int}.

Matrix multiplication is a fundamental building block of many science and engineering fields, such as machine learning, image and signal processing, wireless communication, optimization and so on. In this paper, we focus on the problem of secure distributed matrix multiplication. Secure matrix multiplication has been studied in cryptography community, and different approaches have been proposed, including a weaker version of fully homomorphic encryption, namely partially homomorphic encryption \cite{SMM2013Khan,SMM2017Xavier,SMM2016Duong}.

In contrast to the focus of cryptography community, there are not many works on secure matrix multiplication using information theoretic tools. A lot of efforts are put in further speeding up computation and reducing communication overhead using codes when it comes to distributed matrix multiplication in information theory community. Several recent works include \cite{PolyCode2017,PolyCode2018,MatDot}. These works speed up matrix multiplication and reduce communication overhead by adding redundancy to the computation using codes. The authors showed that the added redundancy allows the distributed system to tolerate servers who do not respond in a timely manner and mitigate stragglers, and allows the user and servers to communicate less.

\begin{figure*}[t]
\centering
	\includegraphics[width=0.8\linewidth]{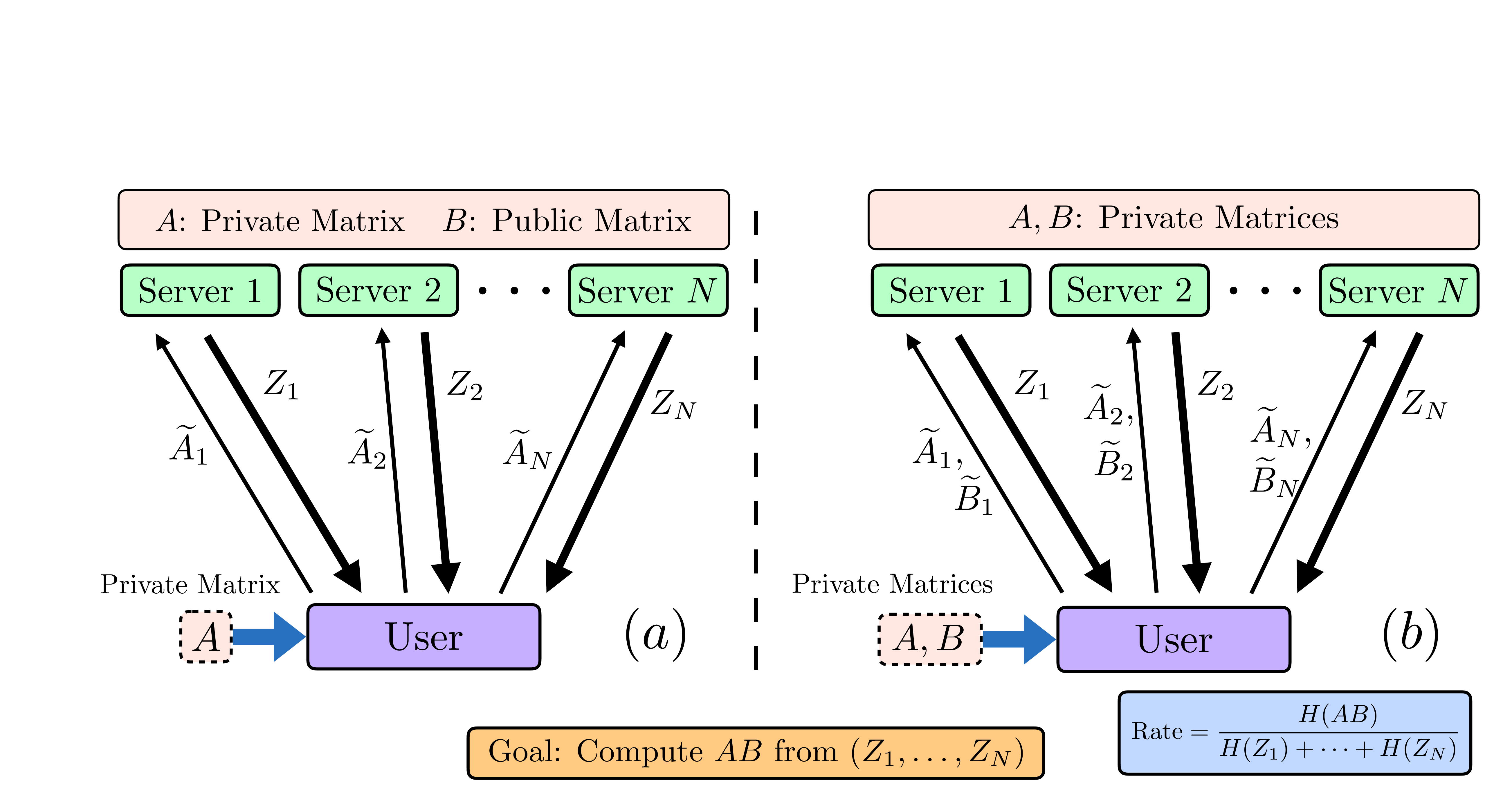}
	\caption{$(a)$ One-sided secure matrix multiplication. $(b)$ Fully secure matrix multiplication.
	\label{Fig:model}}
	\vspace{-15pt}
\end{figure*} 

\textit{Main Contributions:} In this work, we wish to combine the desirable features of works in both communities, and devise schemes that are both $(a)$ information-theoretically secure; and $(b)$ have the smallest communication overhead. We consider a system including one user connected to $N$ servers. We assume that servers are honest, but curious. The user wishes to multiply $A\in\mathbb{F}^{m\times n}$ and $B\in\mathbb{F}^{n\times p}$. We consider this problem under two different models.
\begin{itemize}
\item We first study the model where $B$ is a public matrix available at all servers, and $A$ is private. The goal is to compute $AB$ securely when any $\ell$ servers may collude. We devise a capacity achieving scheme based on Shamir's secret sharing scheme \cite{Shamir}. We derive an information-theoretic converse proof and show that the capacity is $(N-\ell)/N$. This result shows that for a scheme to be secured against any $\ell$ colluding servers, the price we have to pay is $\ell/N$.
\item We next study the model where both $A$ and $B$ are private matrices with the same goal when any $\ell$ servers can collude. We devise a novel achievable scheme inspired by the recent works, \cite{PolyCode2017,PolyCode2018}, which show how to leverage codes for distributed matrix multiplication. For this model, our scheme achieves a rate of $(\lceil \sqrt{N}-\ell \rceil)^2/(\lceil \sqrt{N}-\ell \rceil+\ell)^2$. We also show that there is room for improvement and provide an example of the improved scheme.
\end{itemize}

The rest of the paper is organized as follows. We describe the system and problem statement in Section \ref{Sec:SystemModel}. We present the capacity achieving scheme and the converse proof for the one-sided secure matrix multiplication in Section \ref{Sec:Collude}, and the achievable scheme for the fully secure matrix multiplication in Section \ref{Sec:FullySecure}. We conclude the paper in Section \ref{Sec:Conclusion}.

\section{System Model and Problem Formulation \label{Sec:SystemModel}}
We consider a problem where there are $N$ servers, and a user who wants to compute the product of two input matrices $A\in\mathbb{F}^{m\times n}$ and $B\in\mathbb{F}^{n\times p}$ securely, i.e., $AB$, using $N$ servers, for some integer $m,n$ and $p$, and a sufficiently large field $\mathbb{F}$. The user is connected to each server through a private link (see Fig. \ref{Fig:model}) and we assume that the servers are honest, but curious. In order to prevent servers from learning about input matrices, the user sends securely encoded versions of input matrices to servers. We define the encoding functions as,
\begin{align}
\bm{f}=(\mathit{f}_1,\mathit{f}_2,\dots,\mathit{f}_{N}),\quad \bm{g}=(\mathit{g}_1,\mathit{g}_2,\dots,\mathit{g}_{N}),
\end{align}
where $\mathit{f}_i$ and $\mathit{g}_i$ are the encoding functions for server $i$. The encoded matrices for server $i$ are denoted by $\widetilde{A}_i$ and $\widetilde{B}_i$ for two input matrices for $i=1,2,\dots,N$, i.e.,
\begin{align}
\widetilde{A}_i=\mathit{f}_i(A),\quad \widetilde{B}_i=\mathit{g}_i(B).
\end{align}
The dimensions of $\widetilde{A}_i$ and $\widetilde{B}_i$ vary depending on the scheme used. We denote the answer from server $i$ as $Z_i,i=1,2,\dots,N$. From all answers $Z_1,Z_2,\dots,Z_{N}$, the user must be able to decode the desired result $AB$. We define the decoding function as $d(.)$, therefore, $AB=d(Z_1,Z_2,\dots,Z_{N})$. Hence, decodability constraint can be written as,
\begin{align}
H(AB\vert Z_1,Z_2,\dots,Z_{N})=0 \label{Reliable}.
\end{align}

In this paper, we study the following two models:

\noindent $(a)$ \textit{One-Sided Secure Matrix Multiplication with $\ell$ Colluding Servers:} In this model, $B$ is a public arbitrary constant matrix available at all servers, where $A$ is a private random matrix at the user. Our goal is to securely multiply $A$ and $B$ without revealing anything about $A$ even when any $\ell$ servers may collude (see Fig. \ref{Fig:model}($a$)), i.e., colluding servers can gather their respective received matrix $\widetilde{A}_i$ and attempt to learn about $A$. The user does not know which $\ell$ servers may collude. We use the index set $\mathcal{L}=\{i_1, i_2,\dots,i_{\ell}\}\subseteq[1:N],\vert\mathcal{L}\vert=\ell$ to denote a subset of $\ell$ servers, and $\widetilde{A}_\mathcal{L}\triangleq (\widetilde{A}_{i_1},\widetilde{A}_{i_2},\dots,\widetilde{A}_{i_{\ell}})$ to denote the corresponding encoded version of A sent to servers in the set $\mathcal{L}$. For a scheme in this setting to be considered secured, the encoded matrices $\widetilde{A}_\mathcal{L},\forall \mathcal{L}\subseteq[1:N], \vert\mathcal{L}\vert=\ell$ must not leak anything about $A$. Thus, a scheme for this model must satisfy the following security constraint,
\begin{align}
I(A;\widetilde{A}_{\mathcal{L}})=0,\forall \mathcal{L}\subseteq[1:N], \vert\mathcal{L}\vert=\ell \label{Constraint:Secure2}.
\end{align}

We say that the rate $R$ is achievable if there exists a scheme satisfying the decodability and security constraints, i.e., (\ref{Reliable}) and (\ref{Constraint:Secure2}). The rate is characterized by the number of desired bits per download bit. The rate is defined as,
\begin{align}
R=\frac{H(AB)}{\sum\limits_{i=1}^{N} H(Z_i)}.
\end{align}
The capacity $C_{\text{one-sided}}^{(\ell)}$ is the supremum of $R$ over all feasible schemes for this model.

\noindent $(b)$ \textit{Fully Secure Matrix Multiplication with $\ell$ Colluding Servers:} In this model, both $A$ and $B$ are private matrices at the user. Our goal is to multiply them securely when any $\ell$ servers may collude (see Fig. \ref{Fig:model}($b$)). Hence, encoded matrices $\widetilde{A}_\mathcal{L}$ and $\widetilde{B}_\mathcal{L},\forall \mathcal{L}\subseteq[1:N], \vert\mathcal{L}\vert=\ell$ must not reveal anything about $A$ and $B$. The security constraint for this model is,
\begin{align}
I(A,B;\widetilde{A}_\mathcal{L},\widetilde{B}_\mathcal{L})=0,\forall \mathcal{L}\subseteq[1:N], \vert\mathcal{L}\vert=\ell \label{Constraint:Secure3}.
\end{align}

We say that the rate $R$ is achievable if there exits a scheme for which it satisfies both (\ref{Reliable}) and (\ref{Constraint:Secure3}). Similarly, $C_{\text{fully}}^{(\ell)}$ is defined as the supremum of achievable rates for the fully secure matrix multiplication problem. It is clear that $C_{\text{one-sided}}^{(\ell)}\geq C_{\text{fully}}^{(\ell)}$. In the next two sections, we present our main results towards characterizing these capacities.

\section{One-Sided Secure Matrix Multiplication with $\ell$ Colluding Servers \label{Sec:Collude}}
We first study the model where $B$ is public and known at all servers, and the user wants to securely compute $AB$ without revealing $A$ to any $\ell$ colluding servers. We present our proposed scheme, followed by a converse proof to show that the scheme is information-theoretically optimal.
\begin{theorem}
For the $(N,\ell)$ one-sided secure matrix multiplication problem, in which $B$ is known everywhere and $A$ is kept hidden from any $\ell$ colluding servers while computing $AB$, the capacity is given by
\label{Theorem2}
\end{theorem}
\vspace{-15pt}
\begin{align}
C_{\text{one-sided}}^{(\ell)}=\frac{N-\ell}{N}.
\end{align}
Before presenting the achievable scheme, we first show an example to highlight the intuition behind the scheme.

\begin{example}
$(N=4,\ell=2)$ Consider a one-sided secure matrix multiplication problem with $4$ servers, and any $2$ of them can collude. The user partitions $A$ into
\begin{align}
A=
\begin{bmatrix}
A_1\\
A_2
\end{bmatrix},
\end{align}
where $A_1,A_2\in\mathbb{F}^{(m/2)\times n}$. The original matrix multiplication can be rewritten as,
\begin{align}
AB=
\begin{bmatrix}
A_1B\\
A_2B
\end{bmatrix}.
\end{align}
The goal is now to recover $A_1B$ and $A_2B$. The user generates $2$ random matrices, i.e., $K_1,K_2\in\mathbb{F}^{(m/2)\times n}$, whose entries are i.i.d. uniform random variables from the field $\mathbb{F}$, and encodes the matrix for server $i$ as,
\begin{align}
\widetilde{A}_i&=A_1+iA_2+i^2K_1+i^3K_2,
\end{align}
where each $\widetilde{A}_i$ has the same dimension as $A_1$ and $A_2$ for all $i=1,2,3,4$. Server $i$ computes $\widetilde{A}_iB$ and returns the result to the user. The results received at the user are,
\begin{align}
Z_1&=\widetilde{A}_1B=A_1B+A_2B+K_1B+K_2B,\nonumber\\
Z_2&=\widetilde{A}_2B=A_1B+2A_2B+4K_1B+8K_2B,\nonumber\\
Z_3&=\widetilde{A}_3B=A_1B+3A_2B+9K_1B+27K_2B,\nonumber\\
Z_4&=\widetilde{A}_4B=A_1B+4A_2B+16K_1B+64K_2B.
\end{align}
Clearly, the results can be viewed as a system of $4$ equations in $4$ matrices, and rewritten in matrix form as,
\begin{align}
\begin{bmatrix}
Z_1\\
Z_2\\
Z_3\\
Z_4
\end{bmatrix}
=
\begin{bmatrix}
1^0 & 1^1 & 1^2 & 1^3\\
2^0 & 2^1 & 2^2 & 2^3\\
3^0 & 3^1 & 3^2 & 3^3\\
4^0 & 4^1 & 4^2 & 4^3
\end{bmatrix}
\begin{bmatrix}
A_1B\\
A_2B\\
K_1B\\
K_2B
\end{bmatrix}.
\end{align}
Since the coefficient matrix is a Vandermonde matrix, the system is invertible with a unique solution. The user can multiply the inverse of the coefficient matrix on both sides and solve for $A_1B$ and $A_2B$. However, for any $2$ servers, they see a system of $2$ equations in $4$ matrices, hence, they will not be able to solve for $A_1B$ and $A_2B$. The user is able to recover $2$ desired items from a total of $4$ items, hence, achieving a rate of $1/2$.
\end{example}

\noindent \textit{Proof of Theorem \ref{Theorem2}}

We next present the generalized achievable scheme. We show that the capacity can be achieved by a modified Shamir's secret sharing scheme, and we derive an information-theoretic converse proof for optimality.

\subsection{Achievable Scheme \label{Sec:OneSidedScheme}}
For the achievable scheme, the user first divides $A$ into $N-\ell$ submatrices vertically, i.e.,
\begin{align}
A=[A_1~A_2\dots A_{N-\ell}]^T,
\end{align}
where $A_i\in\mathbb{F}^{(m/(N-\ell))\times n},i=1,2,\dots,N-\ell$. Then the matrix multiplication can be written as
\begin{align}
AB=[A_1B~\dots~A_{N-\ell}B]^T.
\end{align}
The goal is to recover $A_1B,\dots,A_{N-\ell}B$. The user then encodes the submatrices of $A$ into the following form,
\begin{align}
\widetilde{A}_i=\sum\limits_{j=1}^{N-\ell}A_j x_i^{j-1}+\sum\limits_{k=1}^{\ell}K_k x_i^{k+(N-\ell)-1},\label{EncodedAi}
\end{align}
where the dimension of $\widetilde{A}_i$ is the same as any $A_i$, and $x_i$ is a distinct non-zero element in $\mathbb{F}$ assigned to server $i$. All random matrices, $K_1,\dots,K_{\ell}\in\mathbb{F}^{(m/(N-\ell))\times n}$, are generated i.i.d. uniformly at random. (\ref{EncodedAi}) can be seen as a polynomial evaluated at point $x_i$. Servers then multiply their received $\widetilde{A}_i$'s with B and return the following polynomial,
\begin{align}
h(x)=\sum\limits_{j=1}^{N-\ell}A_jB x^{j-1}+\sum\limits_{k=1}^{\ell}K_kB x^{k+(N-\ell)-1},
\end{align}
at $x=x_i,i=1,\dots,N$. Recall that the goal is to recover $A_1B,\dots,A_{N-\ell}B$ from all $Z_i$, i.e., $h(x_i),i=1,\dots,N$. As shown in the example, due to the design of the scheme, the answers can be seen as a system of $N$ equations in $N$ matrices. Since the coefficient matrix is a Vandermonde matrix, the user can multiply the inverse of the coefficient matrix and solve for the desired items. However, a more efficient decoding method is to view each answer $Z_i$ as a degree $N-1$ polynomial evaluated at point $x_i$. The coefficients of a degree $N-1$ polynomial can be recovered with $N$ evaluations by polynomial interpolation. Since we can recover $N-\ell$ desired items from $N$ answers, we achieve a rate of $(N-\ell)/N$.

We next prove that the scheme is secure, i.e., the security constraint (\ref{Constraint:Secure2}) is satisfied. We start from the following sequence of inequalities:
\begin{align}
I(A;\widetilde{A}_\mathcal{L})\nonumber&= I(A;\widetilde{A}_{i_1},\dots,\widetilde{A}_{i_{\ell}})\nonumber\\
&= H(\widetilde{A}_{i_1},\dots,\widetilde{A}_{i_{\ell}}) - H(\widetilde{A}_{i_1},\dots,\widetilde{A}_{i_{\ell}}\vert A)\nonumber\\
&\stackrel{(a)}{=} H(\widetilde{A}_{i_1},\dots,\widetilde{A}_{i_{\ell}}) - H(K_1,\dots,K_{\ell})\nonumber\\
&\stackrel{(b)}{=} H(\widetilde{A}_{i_1},\dots,\widetilde{A}_{i_{\ell}}) - \ell\frac{mn}{N-\ell}\log\vert\mathbb{F}\vert\nonumber\\
&\stackrel{(c)}{\leq} H(\widetilde{A}_{i_1})+\dots+H(\widetilde{A}_{i_{\ell}}) - \ell\frac{mn}{N-\ell}\log\vert\mathbb{F}\vert\nonumber\\
&\stackrel{(d)}{=} \ell\frac{mn}{N-\ell}\log\vert\mathbb{F}\vert - \ell\frac{mn}{N-\ell}\log\vert\mathbb{F}\vert=0,\label{SecureProof}
\end{align}
where $(a)$ follows from (\ref{EncodedAi}) and the fact that all random matrices $K_j$'s are independent of $A$, and $(b)$ due to the entropy of a uniformly distributed random variable being $\log\vert\mathbb{F}\vert$ and the dimension of each one of the $\ell$ random matrices $K_j$ being $mn/(N-\ell)$, $(c)$ follows by upper bounding the joint entropy using the sum of individual entropies and $(d)$ follows from the argument similar to $(b)$. Since mutual information is non-negative and it is upper bounded by zero, we conclude that the scheme is information-theoretically secure.
\subsection{Converse}
We start the converse proof from the following sequence of inequalities:
\begin{align}
H(AB)&= H(AB)-H(AB\vert Z_1,\dots,Z_{N})+\underbrace{H(AB\vert Z_1,\dots,Z_{N})}_{=0}\nonumber\\
&\stackrel{(a)}{=} I(AB;Z_1,\dots Z_{N})\nonumber\\
&= H(Z_1,\dots,Z_{N})-H(Z_1,\dots,Z_{N}\vert AB)\nonumber\\
&\stackrel{(b)}{\leq} H(Z_1,\dots,Z_{N})-H(Z_{i_1},\dots,Z_{i_{\ell}}\vert AB) \nonumber\\
&\stackrel{(c)}{=} H(Z_1,\dots,Z_{N})-H(Z_\mathcal{L}),\label{ConversePart1}
\end{align}
where $(a)$ is due to decodability constraint (\ref{Reliable}), $(b)$ follows by lower bounding the joint entropy of $N$ items using the joint entropy of $\ell$ items, $(c)$ follows from the Markov Chain $A\rightarrow\widetilde{A}_\mathcal{L}\rightarrow Z_\mathcal{L}$ and the fact that from data-processing inequality, we know $I(A;\widetilde{A}_\mathcal{L})\geq I(A;Z_\mathcal{L})$, which is greater than $I(AB;Z_\mathcal{L})$. This indicates that $I(AB;Z_\mathcal{L})=0$, hence, we get $H(Z_\mathcal{L}\vert AB)=H(Z_\mathcal{L}),\mathcal{L}\subseteq\{1,\dots,N\},\vert\mathcal{L}\vert=\ell$. Since there are ${N}\choose{\ell}$ possible subsets $\mathcal{L}$ of servers of size $\ell$, we sum up their entropy and have,
\begin{align}
{{N}\choose{\ell}}H(AB)\leq {{N}\choose{\ell}}H(Z_1,\dots,Z_{N})-\sum\limits_{\substack{\vert\mathcal{L}\vert=\ell\\ \mathcal{L}\subseteq\{ 1,\dots,N \}}} H(Z_\mathcal{L}).\label{ConversePart2}
\end{align}
Rearranging (\ref{ConversePart2}), we have,
\begin{align}
H(AB) &\leq H(Z_1,\dots,Z_{N})
-\ell\frac{1}{{{N}\choose{\ell}}}\sum\limits_{\substack{\vert\mathcal{L}\vert=\ell\\ \mathcal{L}\subseteq\{ 1,\dots,N \}}} \frac{H(Z_\mathcal{L})}{\ell}\nonumber\\
&\stackrel{(a)}{\leq} H(Z_1,\dots,Z_{N})-\ell\frac{H(Z_1,\dots,Z_{N})}{N}\nonumber\\
&=\left(1-\frac{\ell}{N}\right)H(Z_1,\dots,Z_{N})\nonumber\\
&\stackrel{(b)}{\leq} \left(\frac{N-\ell}{N}\right)\sum\limits_{i=1}^{N} H(Z_i),\label{ConversePart3}
\end{align}
where in $(a)$ we apply Han's inequality \cite[Chapter $17$]{ThomasCover} to bound the second term and $(b)$ follows by bounding the joint entropy using the sum of entropies. From (\ref{ConversePart3}), we get
\begin{align}
R_\text{one-sided}^{(\ell)}=\frac{H(AB)}{\sum\limits_{i=1}^{N} H(Z_i)}\leq \frac{N-\ell}{N}.\label{OneSidedRate}
\end{align}
Hence, from the upper bound in (\ref{OneSidedRate}) and a matching scheme in Section \ref{Sec:OneSidedScheme}, we conclude that the capacity for the one-sided matrix multiplication problem is $C_{\text{one-sided}}^{(\ell)}=(N-\ell)/N$. This completes the proof of Theorem \ref{Theorem2}.

\section{Fully Secure Matrix Multiplication with $\ell$ Colluding Servers \label{Sec:FullySecure}}
We next investigate the case where the user wants to compute $AB$ securely while keeping $A$ and $B$ information-theoretically secure from any $\ell$ colluding servers. We next present our main result for the fully secure matrix multiplication in the following Theorem.
\begin{theorem}
For the $(N,\ell)$ fully secure matrix multiplication problem, in which both $A$ and $B$ must be kept secure from any $\ell$ colluding servers while computing $AB$, we have the following lower bound on the capacity:
\label{Theorem3}
\end{theorem}
\vspace{-15pt}
\begin{align}
C_\text{fully}^{(\ell)}\geq \frac{(\ceiling{\sqrt{N}-\ell})^2}{(\ceiling{\sqrt{N}-\ell}+\ell)^2}\label{RateFully}.
\end{align}

\begin{figure}[t]
\centering
	\includegraphics[width=0.50\linewidth]{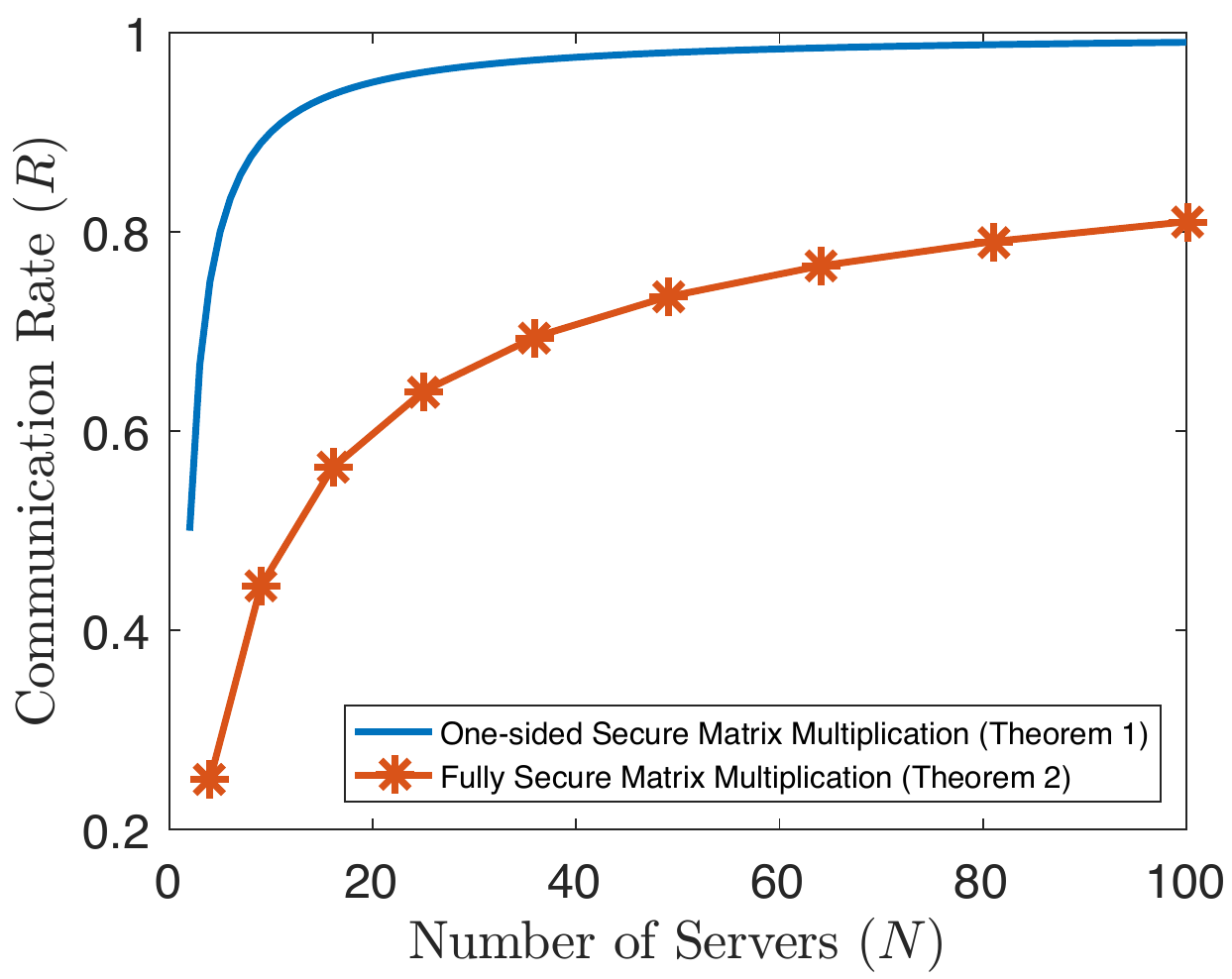}
	\caption{Comparison between the communication rates of One-sided and Fully secure schemes for $\ell=1$ as $N$ is varied.
    \label{Fig:SMML1}}
	\vspace{-15pt}
\end{figure}

\begin{figure}[t]
\centering
	\includegraphics[width=0.50\linewidth]{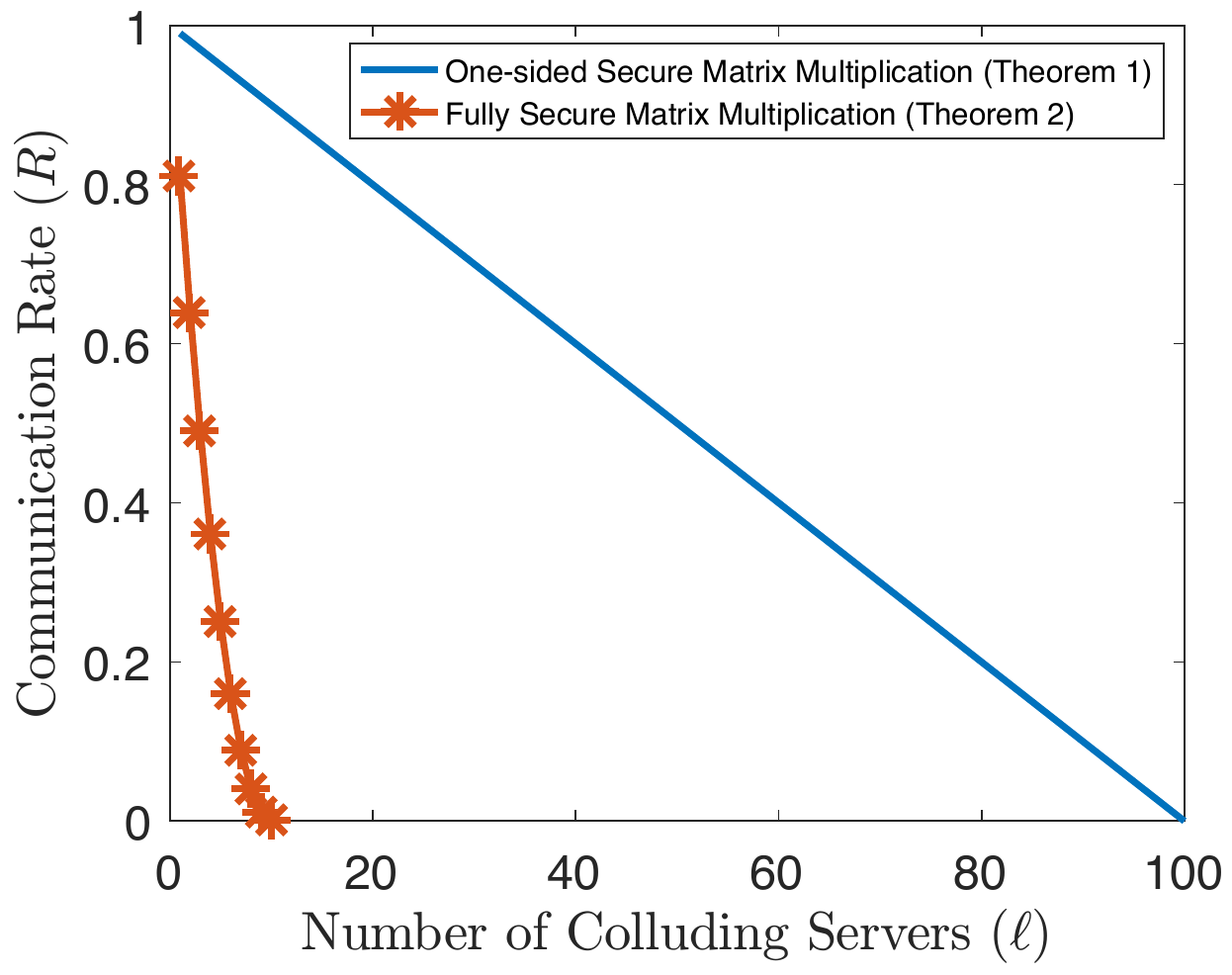}
	\caption{The impact of number of colluding servers on the communication rate when $N=100$.
    \label{Fig:SMMN100}}
	\vspace{-16pt}
\end{figure}

Before presenting the proposed scheme, we first compare the achievable rate of the proposed fully secure scheme to the capacity of the one-sided secure matrix multiplication problem. Clearly, due to a stronger security requirement, it is clear that the rate of the proposed fully secure scheme to be lower than the capacity of the one-sided secure matrix multiplication problem, when the number of colluding servers $\ell$ is fixed at a certain value. In Fig. \ref{Fig:SMML1}, we let $\ell=1$ and increase the number of total servers $N$. It can be seen that the rate of the the proposed scheme of Theorem \ref{Theorem3} is lower, compare to the rate of Theorem \ref{Theorem2}. Notably, both schemes converge to $1$ asymptotically as $N\rightarrow\infty$, however, the convergence for Theorem \ref{Theorem2} is faster than the convergence for Theorem \ref{Theorem3}. We can also see from Fig. \ref{Fig:SMMN100}, that the rate of the proposed scheme decreases a lot faster than the capacity of the one-sided secure matrix multiplication problem when $N$ is fixed to $100$ and $\ell$ is changing. This indicates that our proposed scheme cannot tolerate too many colluding servers due to the $\sqrt{N}$ term in (\ref{RateFully}). We next present the proposed scheme in detail.

\subsection{Proof of Theorem \ref{Theorem3} \label{Sec:ProposedScheme}}
For the $(N,\ell)$ fully secure matrix multiplication problem, the user wishes to compute $AB$ securely without revealing either $A$ or $B$ when any $\ell$ servers may collude. The user breaks the input matrices into $r$ submatrices, where $r=\ceiling{\sqrt{N}-\ell}$. The reason for choosing this value of $r$ will become clear when we fully describe the scheme next. The submatrices are,
\begin{align}
A=
\begin{bmatrix}
A_1\\
A_2\\
\vdots\\
A_{r}
\end{bmatrix}
\text{and}
~B=
\begin{bmatrix}
B_1~B_2 \dots B_{r}
\end{bmatrix},
\end{align}
where $A_i\in\mathbb{F}^{(m/r)\times n}$ and $B_i\in\mathbb{F}^{n\times (p/r)},\forall i$. Hence, we write the matrix multiplication $AB$ as,
\begin{align}
\begin{bmatrix}
A_1\\
A_2\\
\vdots\\
A_{r}
\end{bmatrix}
\begin{bmatrix}
B_1~B_2 \dots B_{r}
\end{bmatrix}
=
\begin{bmatrix}
A_1B_1 & A_1B_2 & \dots & A_1B_r\\
A_2B_1 & A_2B_2 & \dots & A_2B_r\\
\vdots & \vdots & \ddots & \vdots\\
A_rB_1 & A_rB_2 & \dots & A_rB_r
\end{bmatrix}
,
\end{align}
where the original matrix multiplication can be seen as composed of $r^2$ smaller matrix multiplications.

Similar to one-sided secure matrix multiplication problem, the user generates $\ell$ random matrices $K_{A_1},\dots,K_{A_{\ell}}\in\mathbb{F}^{(m/r)\times n}$ for $A$, and $\ell$ random matrices $K_{B_1},\dots,K_{B_{\ell}}\in\mathbb{F}^{n\times (p/r)}$ for $B$, where each of their entries is an i.i.d. uniform random variable. The user encodes $A$ and $B$ for server $i$ as follows:
\begin{align}
\widetilde{A}_i&=\sum\limits_{j=1}^{r}A_jx_i^{j-1}+\sum\limits_{k=1}^{\ell}K_{A_k}x_i^{k+r-1}\label{FullyAi},\\
\widetilde{B}_i&=\sum\limits_{j=1}^{r}B_jx_i^{(j-1)(r+\ell)}+\sum\limits_{k=1}^{\ell}K_{B_k}x_i^{(k+r-1)(r+\ell)}\label{FullyBi},
\end{align}
where $\widetilde{A}_i\in\mathbb{F}^{(m/r)\times n}$ and $\widetilde{B}_i\in\mathbb{F}^{n\times (p/r)}$. The degrees of (\ref{FullyAi}) and (\ref{FullyBi}) are chosen in a way that each item is guaranteed to be the only item at a certain degree after multiplication. This methodology is similar to the one proposed in \cite{PolyCode2017,PolyCode2018} for distributed matrix multiplication problem. Essentially, computing $\widetilde{A}_i\widetilde{B}_i$ is equivalent to evaluating the following polynomial with $4$ different types of terms:
\newpage
\begin{align}
\mathit{h}(x)&=\sum\limits_{j=1}^{r}\sum\limits_{j'=1}^{r}A_jB_{j'}x^{j+(j'-1)(r+\ell)-1}\nonumber\\
&+\sum\limits_{j=1}^{r}\sum\limits_{k'=1}^{\ell}A_jK_{B_{k'}}x^{j+(k'+r-1)(r+\ell)-1}\nonumber\\
&+\sum\limits_{k=1}^{\ell}\sum\limits_{j'=1}^{r}K_{A_k}B_{j'}x^{k+r+(j'-1)(r+\ell)-1}\nonumber\\
&+\sum\limits_{k=1}^{\ell}\sum\limits_{k'=1}^{\ell}K_{A_k}K_{B_{k'}}x^{k+r+(k'+r-1)(r+\ell)-1}.\label{AnswerI}
\end{align}

Due to the design of the scheme, each degree has exactly one item as its coefficient in (\ref{AnswerI}). Note that the polynomial has degree $(r+\ell)^2-1$, hence, evaluations at $(r+\ell)^2$ distinct points are sufficient to solve for all coefficients of the polynomial. This indicates that we need at least $(r+\ell)^2$ responses, one from each server to recover the desired result, i.e., $N\geq(r+\ell)^2$. However, the user is only interested in the first double summation term in (\ref{AnswerI}), which has a total of $r^2$ items in the form of $A_jB_{j'}$. Since the user can recover $r^2$ items out of $(r+\ell)^2$ items, the achievable scheme yields a rate of $r^2/(r+\ell)^2=(\ceiling{\sqrt{N}-\ell})^2/(\ceiling{\sqrt{N}-\ell}+\ell)^2$.

We next show that the proposed scheme is information-theoretically secure:
\begin{align}
&~\quad I(A,B;\widetilde{A}_\mathcal{L},\widetilde{B}_\mathcal{L})\nonumber\\
&= I(A,B;\widetilde{A}_\mathcal{L})+I(A,B;\widetilde{B}_\mathcal{L}\vert \widetilde{A}_\mathcal{L})\nonumber\\
&= H(\widetilde{A}_\mathcal{L})-H(\widetilde{A}_\mathcal{L}\vert A,B)+ H(\mset{B}\vert\mset{A})-H(\mset{B}\vert\mset{A},A,B)\nonumber\\
&\stackrel{(a)}{=} H(\mset{A})-H(K_{A_1},\dots,K_{A_\ell})+H(\mset{B})-H(K_{B_1},\dots,K_{B_\ell})\nonumber\\
&\stackrel{(b)}{=} H(\mset{A})-\ell\frac{mn}{r}\log\vert\mathbb{F}\vert+H(\mset{B})-\ell\frac{np}{r}\log\vert\mathbb{F}\vert\nonumber\\
&\stackrel{(c)}{\leq} H(\widetilde{A}_{i_1})+\dots+H(\widetilde{A}_{i_\ell})-\ell\frac{mn}{r}\log\vert\mathbb{F}\vert+ H(\widetilde{B}_{i_1})+\dots+H(\widetilde{B}_{i_\ell})-\ell\frac{np}{r}\log\vert\mathbb{F}\vert\nonumber\\
&\stackrel{(d)}{=} \ell\frac{mn}{r}\log\vert\mathbb{F}\vert - \ell\frac{mn}{r}\log\vert\mathbb{F}\vert + \ell\frac{np}{r}\log\vert\mathbb{F}\vert - \ell\frac{np}{r}\log\vert\mathbb{F}\vert\nonumber\\
&= 0, \label{FullySecure1}
\end{align}
where $(a)$ follows from (\ref{FullyAi}), (\ref{FullyBi}) and the fact that random matrices are independent of $A$ and $B$, and $\widetilde{B}_\mathcal{L}$ is independent of $\widetilde{A}_\mathcal{L}$, $(b)$ follows by summing the entropy of each uniformly distributed random variable in all $K_{A_j}$ and $K_{B_{j'}}$, $(c)$ follows by upper bounding the joint entropy using the sum of individual entropies, $(d)$ follows from the argument similar to $(b)$. Hence, the proposed scheme is information-theoretically secure. This completes the proof of Theorem \ref{Theorem3}.

\subsection{Improving Theorem \ref{Theorem3} by \textit{Aligned} Secret Sharing}
Due to the design of our proposed scheme, each item is the coefficient of a distinct degree. However, in a fully secure matrix multiplication problem, only items with the form of $A_jB_{j'}$ are useful. Hence, if we can ensure that each item with the form of $A_jB_{j'}$ is the only coefficient of some distinct degrees while aligning the other undesired items, potentially, we can achieve a better rate. We present the following example to demonstrate the idea of aligned secret sharing.

\begin{example}
Consider the $(8,1)$ fully secure matrix multiplication problem where there are $8$ servers, and none of them collude. For this example, from Theorem \ref{Theorem3}, we can achieve a rate of $(\ceiling{\sqrt{N}-\ell})^2/(\ceiling{\sqrt{N}-\ell}+\ell)^2=2^2/(2+1)^2=4/9$. We now show how to improve upon this rate through the aligned secret sharing scheme.

The user partitions $A$ and $B$ into the following
\begin{align}
A=
\begin{bmatrix}
A_1\\
A_2
\end{bmatrix}
\text{ and }
B=
\begin{bmatrix}
B_1 & B_2
\end{bmatrix},
\end{align}
where $A_1,A_2\in\mathbb{F}^{(m/2)\times n}$, and $B_1,B_2\in\mathbb{F}^{n\times (p/2)}$. The user generates one random matrix for each $A$ and $B$, i.e., $K_A\in\mathbb{F}^{(m/2)\times n}$ and $K_B\in\mathbb{F}^{n\times (p/2)}$. Instead of following the proposed scheme in Section \ref{Sec:ProposedScheme}, we align the undesired terms in the forms of $A_jK_B,K_AB_{j'}$ and $K_AK_B$ by selecting different degrees for the encoding polynomial. For each server, the encoding of the user is:
\begin{align}
\widetilde{A}_i&=A_1+A_2x_i+K_{A}x_i^2\\
\widetilde{B}_i&=B_1+B_2x_i^3+K_{B}x_i^5,
\end{align}
where $\widetilde{A}_i$ and $\widetilde{B}_i$ have the same dimension as $A_i$ and $B_i$ for $i=1,\dots,8$. Each server $i$ evaluates the polynomial
\begin{align}
h(x_i&)=A_1B_1+A_2B_1x_i+K_{A}B_1x_i^2+A_1B_2x_i^3+A_2B_2x_i^4+(K_AB_2+A_1K_B)x_i^5+A_2K_Bx_i^6+K_AK_Bx_i^7,
\end{align}
for $i=1,\dots,8$. Clearly, the desired items are the only coefficients of their respective degrees, consequently, the user can decode them using polynomial interpolation. Since the degree of the polynomial is now $7$, evaluation at $8$ points are sufficient and there are $4$ desired items. The rate is now $4/8=1/2$ which is larger than $4/9$.
\label{Ex:Improved}
\end{example}

\section{Conclusions \label{Sec:Conclusion}}
In this paper, we studied one-sided and fully secure matrix multiplication problems. We proposed a secret sharing based scheme for the one-sided secure matrix multiplication model, where $B$ is a public matrix and $A$ is a private matrix that must not be learned by servers while computing $AB$ when any $\ell$ servers may collude. We completely characterized the capacity for the communication overhead as $(N-\ell)/N$. We also presented a novel achievable scheme for the fully secure matrix multiplication model, where both $A$ and $B$ are private matrices that must not be learned by servers when any $\ell$ of them may collude. We also presented an improvement for this general scheme through the idea of aligned secret sharing. There are several interesting open problems: $(a)$ finding a converse (upper bound) for the fully secure matrix multiplication problem; and $(b)$ general these ideas for other secure distributed computation tasks.


\bibliographystyle{IEEEtran}
\bibliography{Ref}

\end{document}